\documentclass[twocolumn,prx,amsmath,amssymb,superscriptaddress,longbibliography]{revtex4-1}

\usepackage{tikz}
\usepackage{quantikz} 
\usepackage{pgfplots} 
\pgfplotsset{compat = newest} 
\usepgfplotslibrary{ternary} 
\usepackage[T1]{fontenc}
\usepackage{bbold}

\usepackage[citecolor=blue, urlcolor=blue, linkcolor=blue, colorlinks=true]{hyperref}

\definecolor{nice-blue}{HTML}{1F77B4}
\definecolor{nice-green}{HTML}{2CA02C}
\definecolor{nice-orange}{HTML}{FF7F0E}
\definecolor{nice-red}{HTML}{D62728}

\definecolor{blue-1}{HTML}{80caff}
\definecolor{blue-2}{HTML}{4c9ed9}
\definecolor{blue-3}{HTML}{2477b3}
\definecolor{blue-4}{HTML}{0e588c}
\definecolor{blue-5}{HTML}{003c66}

\definecolor{green-1}{HTML}{7ee67e}
\definecolor{green-2}{HTML}{4dbf4d}
\definecolor{green-3}{HTML}{2e992e}
\definecolor{green-4}{HTML}{138013}
\definecolor{green-5}{HTML}{006600}

\definecolor{red-1}{HTML}{ff6666}
\definecolor{red-2}{HTML}{e64545}
\definecolor{red-3}{HTML}{cc2929}
\definecolor{red-4}{HTML}{a61111}
\definecolor{red-5}{HTML}{800000}

\begin{document}
\title{Crossing a topological phase transition with a quantum computer}
\author{Adam Smith}
\email{adam.smith1@nottingham.ac.uk}
\affiliation{Department of Physics, T42, Technische Universit{\"a}t M{\"u}nchen, James-Franck-Stra{\ss}e 1, D-85748 Garching, Germany}
\affiliation{Blackett Laboratory, Imperial College London, London SW7 2AZ, United Kingdom}
\affiliation{School of Physics and Astronomy, University of Nottingham, Nottingham, NG7 2RD, UK}
\author{Bernhard Jobst}
\affiliation{Department of Physics, T42, Technische Universit{\"a}t M{\"u}nchen, James-Franck-Stra{\ss}e 1, D-85748 Garching, Germany}
\author{Andrew G.~Green}
\affiliation{London Centre for Nanotechnology, University College London, Gordon St., London WC1H 0AH, United Kingdom}
\author{Frank Pollmann}
\affiliation{Department of Physics, T42, Technische Universit{\"a}t M{\"u}nchen, James-Franck-Stra{\ss}e 1, D-85748 Garching, Germany}
\affiliation{Munich Center for Quantum Science and Technology (MCQST), Schellingstr. 4, D-80799 M\"unchen, Germany}

\date{\today}

\begin{abstract}
Quantum computers promise to perform computations beyond the reach of modern computers with profound implications for scientific research. Due to remarkable technological advances, small scale devices are now becoming available for use. One of the most apparent applications for such a device is the study of complex many-body quantum systems, where classical computers are unable to deal with the generic exponential complexity of quantum states. Even zero-temperature equilibrium phases of matter and the transitions between them have yet to be fully classified, with topologically protected phases presenting major difficulties. We construct and measure a continuously parametrized family of states crossing a symmetry protected topological phase transition on the IBM Q quantum computers. We present two complementary methods for measuring string order parameters that reveal the transition, and additionally analyse the effects of noise in the device using simple error models. The simulation that we perform is easily scalable and is a practical demonstration of the utility of near-term quantum computers for the study of quantum phases of matter and their transitions.
\end{abstract}

\maketitle

There are now many approaches being taken to realise universal quantum computers~\cite{Nielsen2010}, with numerous academic research groups, companies and governments across the world devoting resources to each. Amongst the most advanced are devices based on trapped ions~\cite{Bruzewicz2019}, localized spins in diamond~\cite{Bradley2019} or silicon~\cite{Yang2019}, and superconducting circuits~\cite{Wendin2017,Kjaergaard2019}. While each has its advantages---such as coherence times, efficient readout, or gate speeds and fidelities---the latter is fast becoming the most adopted approach. Efforts by D-Wave, Google, IBM and Rigetti, for example, all use superconducting circuits based on Josephson junctions. 

Quantum computational technology is still in its infancy, with the state-of-the-art in superconducting qubits consisting of approximately a hundred qubits, 99\% two-qubit gate fidelities, and coherence times of the order of 100$\mu s$~\cite{Kjaergaard2019}. Fault-tolerant error correction is also currently out of reach, and solutions for quantum memory and networking are not fully developed. They are consequently described as Noisy Intermediate-Scale Quantum (NISQ) devices~\cite{Preskill2018}. There are still unanswered questions about the potential utility of NISQ technology and whether there are fundamental obstructions to going beyond this regime. Nevertheless, there has recently been a flurry of proof-of-principle experiments, along with the recent claim of a demonstrable computational advantage using a quantum computer~\cite{Boixo2018,Arute2019}. For example, in the realm of quantum simulation, real quantum devices have been used to find the ground state of small molecules relevant for quantum chemistry~\cite{Kandala2017,OMalley2016}, to measure multi-qubit quantum entanglement~\cite{Choo2018,Wang2018}, and to simulate non-equilibrium quantum dynamics~\cite{Smith2019,Tacchino2019}. This list is far from exhaustive and we do not intend to review the rapid progress of the last decade.

As realised at the very inception of quantum computing~\cite{Feynman1982}, the study of complex many-body quantum systems could benefit tremendously from this new technology. Generically, these systems require the storage and manipulation of an exponentially large number of parameters on a classical computer. By storing and manipulating the quantum state directly on a quantum computer, it may be possible to reach areas of condensed matter physics that are currently intractable. As a relevant example, there does not yet exist a complete classification of topological phases of matter~\cite{Wen2019}. The most interesting and least understood phases occur in two or three dimensions and host exotic non-abelian anyonic quasi-particles~\cite{Nayak2008}, and as a result our most powerful numerical techniques begin to break down. Most notably, quantum Monte Carlo suffers from the sign problem, and dimensionality is a problem for tensor network based methods due to increased entanglement and less efficient contraction schemes when compared with one dimension. On a quantum computer we can avoid classically storing the quantum state, perform sign-problem free computations and work directly with two-dimensional quantum circuits, potentially sidestepping some of the issues plaguing current numerical techniques. This approach has recently been successfully demonstrated using both digital quantum computers~\cite{Satzinger2021} and quantum simulators~\cite{Semeghini2021}.

In this paper we demonstrate how quantum computers can be used to simulate symmetry protected topological phases and the transitions between them.
This paper is structured as follows. In Sec.~\ref{sec: setup} we introduce symmetry protected topological phases and the concrete models we will consider. Using connections to matrix product states, we discuss how we can simulate the ground states of these models exactly in the thermodynamic limit on a quantum computer in Sec.~\ref{sec: infinite}. We then introduce the two methods we use in Sec.~\ref{sec: results}, and show experimental results from the IBM quantum computers~\cite{IBMQ}. In Sec.~\ref{sec: error modelling} we then analyse three simple error models to understand the observed experimental results and the differences between the two methods. And finally we close with a discussion in Sec.~\ref{sec: discussion}.

\section{Setup}\label{sec: setup}
Here, we use the IBM quantum computers to study a symmetry protected topological (SPT) phase of matter~\cite{Pollmann2010,Chen2013}. An SPT phase is one that, as long as certain symmetries are present, is not adiabatically connected to a trivial product state. SPTs cannot be understood in the framework of local order parameters and spontaneous symmetry breaking. Instead they are distinguished by non-local string order parameters~\cite{Haegeman2012,Pollmann2012a,Elben2019}. We consider infinite one dimensional spin-$\frac{1}{2}$ chains described by the three parameter Hamiltonian
\begin{equation}
	\hat{H} = \sum_{i} \left[ -g_{zz}\; \hat{\sigma}^z_i\hat{\sigma}^z_{i+1} - g_x\; \hat{\sigma}^x_i + g_{zxz}\; \hat{\sigma}^z_i\hat{\sigma}^x_{i+1}\hat{\sigma}^z_{i+2} \right].
	\label{eq:H_phase_diagram}
\end{equation}
This Hamiltonian is symmetric under global spin flips generated by $\prod_i \hat{\sigma}^x_i$ as well as time-reversal (complex conjugation). Due to these symmetries the model has a $\mathbb{Z}_2\times \mathbb{Z}_2^T$ SPT phase, as well as a trivial and a symmetry-broken phase. The phase diagram is shown in Fig.~\ref{fig:phase_diagram}~\cite{Verresen2017,Verresen2018}.

We focus on a one dimensional path through this phase diagram, corresponding to the black curve in Fig.~\ref{fig:phase_diagram}, parametrized as $g_{zz} = 2(1-g^2)$, $g_x = (1+g)^2$ and $g_{zxz} = (g-1)^2$, with tuning parameter $g$~\cite{Wolf2006}. This path continuously interpolates between the cluster Hamiltonian $\hat{H}_\text{ZXZ} = 4\sum_i \hat{\sigma}^z_i\hat{\sigma}^x_{i+1}\hat{\sigma}^z_{i+2}$ for $g = -1$ and the trivial paramagnet with Hamiltonian $\hat{H}_\text{X}= -4\sum_i \hat{\sigma}^x_i$ for $g=1$. The transition between the trivial and the SPT phase occurs at the tricritical point between the three phases at $g=0$.

\begin{figure}[t]
	\centering
	\begin{tikzpicture}
		\begin{ternaryaxis}[ternary limits relative=false,
			clip = false,
			xlabel=$g_{zz}$,
			ylabel=$g_{zxz}$,
			zlabel=$g_x$,
			xmin = 0,
			xmax = 4,
			ymin = 0,
			ymax = 4,
			zmin = 0,
			zmax = 4,
			]
		
		    \iftrue
		    \fill[nice-orange!40] (2,2) -- (2,1) -- (0,2) -- (0,4) -- (2,2);
			\fill[nice-green!40] (2,0) -- (2,1) -- (0,2) -- (0,0) -- (2,0);
			\fill[nice-blue!40] (4,0) -- (2,2) -- (2,0) -- (4,0);

			\addplot[no markers, black, ultra thick] table[x = H1, y = H2, z = H3] {phase_diagram.txt};
			
			\node[fill=white,draw,anchor=north west] at (-0.5,4.75) {$\hat{H} \propto \sum_i \hat{\sigma}^z_i\hat{\sigma}^x_{i+1}\hat{\sigma}^z_{i+2}$};
			\node[fill=white,draw,anchor=north east] at (-0.5,-0.25) {$\hat{H} \propto -\sum_i \hat{\sigma}^x_i$};
			\node[fill=white,draw,anchor=south] at (4.433,-0.217) {$\hat{H} \propto -\sum_i \hat{\sigma}^z_i\hat{\sigma}^z_{i+1}$};
			
			\node[fill=white,draw,anchor=north,align=center] at (0.7,2.5) {\footnotesize SPT phase\\[-0.12cm]\footnotesize with cluster state};
			\node[fill=white,draw,anchor=north,align=center] at (0.72,0.75) {\footnotesize Trivial phase\\[-0.12cm]\footnotesize with product state};
			\node[fill=white,draw,anchor=north,align=center] at (2.8,0.625) {\footnotesize Symmetry-broken\\[-0.12cm]\footnotesize Ising phase};
		\end{ternaryaxis}
	\end{tikzpicture}
	\caption{\textbf{Phase diagram for the $\mathbb{Z}_2\times \mathbb{Z}_2^T$ symmetric Hamiltonian in equation \eqref{eq:H_phase_diagram}.} The green phase is the topologically trivial phase containing the paramagnetic product state, the blue phase is the symmetry-broken phase containing the ferromagnetic ground state of the Ising model and the orange phase is the SPT phase containing the cluster state. The black curve corresponds to the one-dimensional path with tuning parameter $g$ described in the main text.}\label{fig:phase_diagram}
\end{figure}

\begin{figure*}[ht]
	\centering
	\begin{minipage}[c]{.3\textwidth}
    \raisebox{47pt}{\textbf{(a)}}
	\hspace{-8pt}\begin{quantikz}[row sep=0.1cm, column sep=0.1cm]
		\lstick{\raisebox{20pt}{$\vdots$}} & \ldots \hspace{3pt} & \gate[style={fill=nice-blue!20},wires=2][0.6cm]{U} & \qw & \qw & \qw & \qw   \\
		\lstick{\ket{0}} & \qw & \qw & \gate[style={fill=nice-blue!20},wires=2][0.6cm]{U} & \qw & \qw & \qw   \\
		\lstick{\ket{0}} & \qw & \qw & & \gate[style={fill=nice-blue!20},wires=2][0.6cm]{U} & \qw & \qw  \\
		\lstick{\ket{0}} & \qw & \qw & \qw & & \ \ldots \qw &  \\
		\lstick{\raisebox{-20pt}{$\vdots$}} & & & & & &   \\
	\end{quantikz}
	\end{minipage}
	\begin{minipage}[c]{.34\textwidth}
	\raisebox{47pt}{\textbf{(b)}}
	\hspace{-8pt}\begin{quantikz}[row sep=0.1cm, column sep=0.1cm]
		\lstick{\ket{0}} & \gate[style={fill=nice-green!20},wires=2][0.6cm]{U_1} & \qw & \qw & \qw & \qw & \meterD[style={fill=nice-red!20}]{} &  \\
		\lstick{\ket{0}} & & \gate[style={fill=nice-blue!20},wires=2][0.6cm]{U} & \qw & \qw & \qw & \qw & \qw  \\
		\lstick{\ket{0}} & \qw & & \gate[style={fill=nice-blue!20},wires=2][0.6cm]{U} & \qw & \qw & \qw & \qw  \\
		\lstick{\ket{0}} & \qw & \qw & & \ \ldots \qw & & \\
		\lstick{$\vdots$} & & & & \ldots\ & \gate[style={fill=nice-blue!20},wires=2][0.6cm]{U} & \qw & \qw \\
		\lstick{\ket{0}} & \qw & \qw & \qw & \qw & \qw & \meterD[style={fill=nice-red!20}]{} & 
	\end{quantikz}
	\end{minipage}
	\begin{minipage}[c]{.3\textwidth}
	\hspace{-5pt}\raisebox{20pt}{\textbf{(U)}}
	\hspace{-10pt}
	\begin{quantikz}[row sep=0.25cm, column sep=0.15cm]
		\lstick{} & \octrl{1} & \ctrl{1} & \gate{X} & \rstick{} \qw\\
		\lstick{\ket{0}} & \gate{W} & \gate{V} & \qw &\rstick{} \qw
	\end{quantikz}\hfill\\
	\vspace{5pt}
	\hspace{10pt}\raisebox{30pt}{\textbf{(U1)}}
	\hspace{-15pt}
	\begin{quantikz}[row sep=0.25cm, column sep=0.15cm]
		\lstick{\ket{0}} & \gate{H}  & \ctrl{1} & \qw & \qw & \qw & \ctrl{1}\gategroup[wires=2,steps
		=1,style={dashed, rounded corners,fill=gray!30}, background]{Only for $g>0$} & \qw & \qw & \rstick{} \qw\\
		\lstick{\ket{0}} & \qw & \targ{} & \gate{R} & \qw & \qw & \control{} & \qw & \qw & \rstick{} \qw
	\end{quantikz}
	\end{minipage}
	\caption{\textbf{Quantum circuit construction of the states}. \textbf{(a)} Iterative construction of the ground state on an infinite chain. \textbf{(b)} Equivalent finite quantum circuit for measuring observables with finite connected support. Red caps indicate that the end qubits are unphysical and should not be measured. \textbf{(U)} and \textbf{(U1)} are the circuits for the two-qubit gates $U$ and $U_1$, respectively. $X$ is the Pauli-X gate, and the single qubit gates $R, W$ and $V$ are specified in Appendix~\ref{ap: circuit decomposition}.}\label{fig:circuit_structure}
\end{figure*}

The non-trivial SPT phase can be distinguished using string order parameters~\cite{Perez-Garcia2008}, which are non-local observables of macroscopic length $l$. In the limit $l\rightarrow \infty$, the string order parameters are non-zero in one of the two phases and zero in the other. The string order parameters that we consider are of the form
\begin{equation}
    \mathcal{S}^{O}(g) =\langle \psi | \hat{O}_i \Bigg(\prod_{j=i+2}^{k-2} \hat{\sigma}^x_j \Bigg) \hat{O}'_k |\psi\rangle
\label{eq:string-order1}
\end{equation}
with $\hat{O}_i = \hat{\sigma}^z_{i}\hat{\sigma}^y_{i+1}$ and $\hat{O}'_k = \hat{\sigma}^y_{k-1}\hat{\sigma}^z_{k}$ defining $\mathcal{S}^{ZY}(g)$, and $\hat{O}_i = \hat{O}'_k = \mathbb{1}$ defining $\mathcal{S}^{\mathbb{1}}$. The length of the string, $l$, is the distance between the first and last Pauli-operator. For $S^{ZY}$ and $S^\mathbb{1}$ the shortest such string lengths are $l=5$ and $l=3$, respectively. Along our path parametrized by $g$, the string order parameter $\mathcal{S}^{ZY}(g)$ (resp. $\mathcal{S}^{\mathbb{1}}(g)$) is zero for $g>0$ ($g<0$) and equal to $4|g|/(1+|g|)^2$ for $g<0$ ($g>0$). The chosen path has the nice property that the string order parameters are independent of the length of the string and correspond exactly to the values obtained in the thermodynamic limit $l \rightarrow \infty$. This property only holds along the black line in Fig.~\ref{fig:phase_diagram} and away from this line we would generically need a macroscopic length $l$ to sharply differentiate the phases.

\begin{figure}[ht]
	\centering
	\includegraphics[width=\columnwidth]{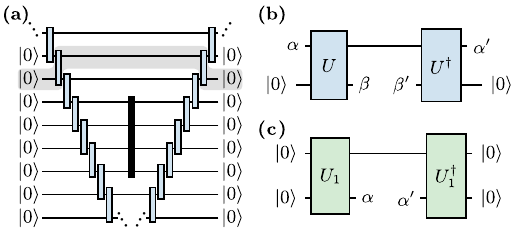}
	\caption{\textbf{Elements of the quantum circuit construction.} \textbf{(a)} Expectation value of an observable is equivalent to sandwiching the operator (black box) between the state and its conjugate. For the infinite system the corresponding circuit contains a repeating element (highlighted in gray). \textbf{(b)} The repeated circuit element away from measured observable when computing expectation values. This circuit element can be interpreted as a transfer matrix, see main text. \textbf{(c)} Corresponding fixed-point vector as a quantum circuit.  }\label{fig:finite-circuit}
\end{figure}

\subsection{Infinite state as finite quantum circuit}\label{sec: infinite}
The ground state of the infinite system can be constructed iteratively by a quantum circuit shown schematically in Fig.~\ref{fig:circuit_structure}(a). We can understand this via a connection to infinite matrix product states~\cite{Vidal2007}, as outlined in Appendix~\ref{ap: MPS}. Any observable with finite connected support can equivalently be measured using the finite quantum circuit in Fig.~\ref{fig:circuit_structure}(b)~\cite{Barratt2020}. That is, any measurement of the qubits---excluding the unphysical first and last qubits---is identical to the corresponding measurement of the infinite chain. In particular, we measure the same energy density $\mathcal{E} = -2(g^2 + 1)$ and values for the string order parameters. Note that this representation of the ground state is exact and in the thermodynamic limit.

We arrive at the finite circuit in Fig.~\ref{fig:circuit_structure}(b) by first viewing a measurement as sandwiching an operator between the quantum circuit (the ket) and the Hermitian conjugate circuit (the bra) as shown in Fig.~\ref{fig:finite-circuit}(a). Away from the observable that we are measuring we find circuit elements of the form shown in Fig.~\ref{fig:finite-circuit}(b). Below the measured operator these will all cancel due to unitarity. While we can't do this for the gates above the measurement, we can construct the gate $U_1$ as a fixed-point of the iterative circuit. More explicitly, we can reinterpret the circuit in Fig.~\ref{fig:finite-circuit}(b) as a transfer matrix $T_{(\beta\beta'),(\alpha\alpha')}$. The expectation value of the operator in the thermodynamic limit is then determined by the fixed points of the transfer matrix, similar to the thermodynamic treatment of the classical 1D Ising model. Similarly, we can consider the circuit in Fig.~\ref{fig:finite-circuit}(c) as a vector $V_{(\alpha\alpha')}$. The unitary $U_1$ is chosen such that $V_{(\alpha\alpha')}$ is the dominant right eigenvector of $T_{(\beta\beta'),(\alpha\alpha')}$ with eigenvalue $1$, i.e. the fixed-point vector under repeated application of the transfer matrix. See Appendix~\ref{ap: MPS} for more details. An alternative way to state the cancellation of the unitary gates below the measurement is that the dominant left eigenvector of the transfer matrix corresponds to the identity.

\section{Results}\label{sec: results}

We will consider two methods for measuring the string order parameters on the IBM quantum computers: \emph{direct measurement}, and \emph{interferometry}. In the former we rotate all of the relevant qubits to the correct basis and then measure all qubit simultaneously. From these measurements we are able to reconstruct the expectation value of Pauli strings, which allows us to measure the energy and string order parameters. In the latter interferometry experiments, we will instead use an additional ancilla qubit which we will entangle with the qubits that we want to measure the string order parameter on. By measuring only this ancilla qubit at the end we can also extract the string order parameter. Details of the interferometry circuits are given in Sec.~\ref{sec: interferometry} and Appendix~\ref{ap: interferometry}. Direct measurement requires a shallower circuit and so is less succeptible to gate errors and decoherence. However, interferometry only requires the measurement of a single qubit in contrast to the direct measurement of many, and so will be impacted less by measurement error. We will compare the accuracy of these two methods due to the competition of different sources of error on current devices.

For our simulations we used the 27 qubit IBM Q device codenamed \texttt{toronto} on 25$^\text{th}$ October 2021, which allows the implementation of a universal gate set consisting of arbitrary single qubit rotations and controlled-not (CNOT) entangling gates between connected qubits. The decomposition of the circuit shown in Fig.~\ref{fig:circuit_structure} into this gate set is given in the methods section. The spins in our system are mapped to the physical qubits of the quantum computer, with the basis states $\{|\! \uparrow \rangle = |0\rangle,  |\! \downarrow\rangle = |1\rangle\}$, and we control the devices using the Python qiskit API~\cite{qiskit}. To select our subset of $N$ qubits we use a custom procedure described in Ref.~\cite{Smith2019}, which maximizes the average CNOT fidelity, while limiting the readout error and coherence time for the qubits. When using the direct method, we also perform error mitigation on the raw data from the machine using methods provided in qiskit~\cite{qiskit}, to reduce the impact of readout errors, which we outline in Appendix~\ref{ap: error mitigation}. We perform 8192 runs for each circuit and omit error bars in our figures since the statistical error is not significant.

\subsection{Direct Measurement}\label{sec: direct}

Figure~\ref{fig:energy} shows the energy density of the state as measured on the IBM device compared with the analytic value, $\mathcal{E} = -2(g^2 + 1)$. We measure the local energies and average over the central qubits excluding the boundary qubits (i.e. $i = 2,\ldots, N-3$), and show the results measured on subsystems of $N=5,\ldots,9$ qubits. Despite the discrepancy in the absolute value, the energy obtained from the quantum computer follows nicely the exact functional form indicating proximity to the target state. The difference between the experimental and the exact result can predominantly be attributed to dephasing and measurment error. The former accounts for the change in shape, with a dip around $g=0$, since it impacts the $\langle\hat{\sigma}^x_j\rangle$ and $\langle\hat{\sigma}^z_{j}\hat{\sigma}^x_{j+1}\hat{\sigma}^z_{j+2}\rangle$ expectation values but not $\langle\hat{\sigma}^z_{j}\hat{\sigma}^z_{j+1}\rangle$. Furthermore, we expect the amount of dephasing to scale with the circuit depth and so with $N$. A demonstration of the effect of dephasing on the energy density is shown in Appendix~\ref{ap: dephasing}. The measurement error on the other hand does not scale with $N$ and accounts for the slight asymmetry between $g<0$ and $g>0$, due to the various lengths of the terms in the Hamiltonian.

\pgfplotsset{every axis legend/.append style={at={(0.5,0.03)},anchor=south,},}
\begin{figure}[t]
	\centering
	\begin{tikzpicture}
		\begin{axis}[
			height = 0.65\columnwidth,
			width = 0.98\columnwidth,
			xtick = {-1,-0.75,-0.5,...,1},
			xmin=-1.08,xmax=1.08,
			ytick = {-1,-1.5,...,-4},
			xlabel = Tuning parameter $g$,
			ylabel = Energy density $\mathcal{E}$,
			grid = major]
		\addplot[color=gray,line width=1.2pt,domain=-1:1,samples=100]{-1*2*(x^2+1)};
		\addlegendentry{exact}
		\addplot[red-1, only marks, mark size=2.2]
		table[x = g, y expr=\thisrowno{1}/1] {data_2021-10-25/real_data_N5_energy_backend_ibmq_toronto_mitigated_2021-10-25.txt};
		\addlegendentry{$N = 5$}
		\addplot[red-2, only marks, mark=square*, mark size=2]
		table[x = g, y expr=\thisrowno{1}/2] {data_2021-10-25/real_data_N6_energy_backend_ibmq_toronto_mitigated_2021-10-25.txt};
		\addlegendentry{$N = 6$}
		\addplot[red-3, only marks, mark=triangle*, mark size=3]
		table[x = g, y expr=\thisrowno{1}/3] {data_2021-10-25/real_data_N7_energy_backend_ibmq_toronto_mitigated_2021-10-25.txt};
		\addlegendentry{$N = 7$}
		\addplot[red-4, only marks, mark=diamond*, mark size=3]
		table[x = g, y expr=\thisrowno{1}/4] {data_2021-10-25/real_data_N8_energy_backend_ibmq_toronto_mitigated_2021-10-25.txt};
		\addlegendentry{$N = 8$}
		\addplot[red-5, only marks, mark=pentagon*, mark size=2.5]
		table[x = g, y expr=\thisrowno{1}/5] {data_2021-10-25/real_data_N9_energy_backend_ibmq_toronto_mitigated_2021-10-25.txt};
		\addlegendentry{$N = 9$}
		\end{axis}
	\end{tikzpicture}
	\caption{\textbf{Average energy density.} We measure the local energy of the state using number of qubits $N=5,6,7,8,9$. The energy is averaged over the central sites excluding the end qubits. The data from the IBM devices is compared with the analytic result.}
	\label{fig:energy}
\end{figure}

Next we show the measurements of the two string order parameters in Fig.~\ref{fig:quant_comp} for lengths $l=5,6,7$ for $\mathcal{S}^{ZY}(g)$, and $l=3,\ldots,7$ for $\mathcal{S}^{\mathbb{1}}(g)$, and compare with the analytic results. Especially for the smallest string lengths, we see qualitative agreement between the results from the quantum computer and the exact results. It appears that for small enough sizes we can well approximate the errors in the device by a constant scaling factor. Importantly, the order parameters are only non-zero in one of the two phases, and tend to zero at the phase transition $g = 0$. However, beyond a certain string length, $l \gtrsim 6$, this information is lost and there is no clear transition point.

As we increase the string length in Fig.~\ref{fig:quant_comp}, the accuracy of the results quickly diminishes, even more so than was observed in Fig.~\ref{fig:energy}. This is due to the fact that we are measuring non-local operators and both the number of qubits and the length of the operator are increasing. For chains of length $N=9$ ($l=7$) we are no longer able to detect the transition, demonstrating the difficulty of constructing and measuring long-range string order in the quantum state due to the current limitations of the quantum computer. Nevertheless, the combination of the measurements of the energy density and the string order parameters confirm that we are able to approximately construct the target states with non-trivial string order on a real quantum computer.

\begin{figure}[t]
	\centering
	\raisebox{145pt}{\textbf{(a)}}
	\hspace{-8pt}
	\begin{tikzpicture}
		\begin{axis}[
			height = 0.65\columnwidth,
			width = 0.98\columnwidth,
			xtick = {-1,-0.75,-0.5,...,1},
			xmin=-1.08,xmax=1.08,
			ytick = {-0.2,-0.1,0,0.1,0.2,0.3,0.4,0.5,0.6,0.7,0.8,0.9,1},
			yticklabels = {-0.2,-0.1,0,0.1,0.2,0.3,0.4,0.5,0.6,0.7,0.8,0.9,1},
			xlabel = Tuning parameter $g$,
			ylabel = String order parameter $\mathcal{S}^{ZY}$,
			grid = major,
			legend pos = north east]
		\addplot[color=gray,line width=1.2pt,domain=-1:0,samples=50]{-1*4*x/(1-x)^2};
		\addlegendentry{exact}
		\addplot[color=gray,line width=1.2pt,domain=0:1,samples=50,forget plot]{0};
		\addplot[blue-3, only marks, mark=triangle*, mark size=3]
		table[x = g, y = sop] {data_2021-10-25/real_data_N7_SO1_backend_ibmq_toronto_mitigated_2021-10-25.txt};
		\addlegendentry{$l = 5$}
		\addplot[blue-4, only marks, mark=diamond*, mark size=3]
		table[x = g, y = sop] {data_2021-10-25/real_data_N8_SO1_backend_ibmq_toronto_mitigated_2021-10-25.txt};
		\addlegendentry{$l=6$}
		\addplot[blue-5, only marks, mark=pentagon*, mark size=2.5]
		table[x = g, y = sop] {data_2021-10-25/real_data_N9_SO1_backend_ibmq_toronto_mitigated_2021-10-25.txt};
		\addlegendentry{$l=7$}
		\end{axis}
	\end{tikzpicture}\\
	\vspace{-3pt}
	\raisebox{145pt}{\textbf{(b)}}
	\hspace{-8pt}
	\begin{tikzpicture}
		\begin{axis}[
			height = 0.65\columnwidth,
			width = 0.98\columnwidth,
			xtick = {-1,-0.75,-0.5,...,1},
			xmin=-1.08,xmax=1.08,
			ytick = {-0.2,-0.1,0,0.1,0.2,0.3,0.4,0.5,0.6,0.7,0.8,0.9,1},
			yticklabels = {-0.2,-0.1,0,0.1,0.2,0.3,0.4,0.5,0.6,0.7,0.8,0.9,1},
			xlabel = Tuning parameter $g$,
			ylabel = String order parameter $\mathcal{S}^\mathbb{1}$,
			grid = major,
			legend pos = north west]
		\addplot[color=gray,line width=1pt,domain=0:1]{1*4*x/(1+x)^2};
		\addlegendentry{exact}
		\addplot[color=gray,line width=1pt,domain=-1:0,forget plot]{0};
		\addplot[blue-1, only marks, mark size=2.2]
		table[x = g, y = sop] {data_2021-10-25/real_data_N5_SO2_backend_ibmq_toronto_mitigated_2021-10-25.txt};
		\addlegendentry{$l=3$}
		\addplot[blue-2, only marks, mark=square*, mark size=2]
		table[x = g, y = sop] {data_2021-10-25/real_data_N6_SO2_backend_ibmq_toronto_mitigated_2021-10-25.txt};
		\addlegendentry{$l=4$}
		\addplot[blue-3, only marks, mark=triangle*, mark size=3]
		table[x = g, y = sop] {data_2021-10-25/real_data_N7_SO2_backend_ibmq_toronto_mitigated_2021-10-25.txt};
		\addlegendentry{$l=5$}
		\addplot[blue-4, only marks, mark=diamond*, mark size=3]
		table[x = g, y = sop] {data_2021-10-25/real_data_N8_SO2_backend_ibmq_toronto_mitigated_2021-10-25.txt};
		\addlegendentry{$l=6$}
		\addplot[blue-5, only marks, mark=pentagon*, mark size=2.5]
		table[x = g, y = sop] {data_2021-10-25/real_data_N9_SO2_backend_ibmq_toronto_mitigated_2021-10-25.txt};
		\addlegendentry{$l=7$}
		\end{axis}
	\end{tikzpicture}
	\caption{\textbf{Identification of the phase transition using direct measurement.} \textbf{(a)} Results of the string order parameter $\mathcal{S}^{ZY}(g)$ of length $l=5,6,7$, simulated using number of qubits $N=7,8,9$, respectively. \textbf{(b)} $\mathcal{S}^{\mathbb{1}}(g)$ of length $l=3,4,5,6,7$, corresponding to $N = 5,6,7,8,9$. We compare with the analytic results.}
	\label{fig:quant_comp}
\end{figure}



\begin{figure}[th!]
    \centering
    \begin{minipage}[c]{\columnwidth}
	\hspace{0pt}\begin{quantikz}[row sep=0.1cm, column sep=0.1cm]
	    \lstick{\ket{0}} & \qw & \qw & \qw & \qw & \qw & \qw & \gate{H} & \ctrl{2} & \gate{H} & \qw & \meter{} \\
		\lstick{\ket{0}} & \gate[style={fill=nice-green!20},wires=2][0.6cm]{U_1} & \qw & \qw & \qw & \qw & \meterD[style={fill=nice-red!20}]{} & & & & & & \\
		\lstick{\ket{0}} & & \gate[style={fill=nice-blue!20},wires=2][0.6cm]{U} & \qw & \qw & \qw & \qw & \qw & \gate[style={fill=nice-blue!20},wires=4][0.6cm]{\hat{S}^O} & \qw & & &\\
		\lstick{\ket{0}} & \qw & & \gate[style={fill=nice-blue!20},wires=2][0.6cm]{U} & \qw & \qw & \qw & \qw & & \qw & & &\\
		\lstick{\ket{0}} & \qw & \qw & & \ \ldots \qw & & & \ldots &\ldots & \!\!\!\!\!\ldots & & &\\
		\lstick{$\vdots$} & & & & \ldots\ & \gate[style={fill=nice-blue!20},wires=2][0.6cm]{U} & \qw & \qw & & \qw & & & \\
		\lstick{\ket{0}} & \qw & \qw & \qw & \qw & \qw &  \meterD[style={fill=nice-red!20}]{} & & & & &
	\end{quantikz}
	\end{minipage}
    \caption{\textbf{Interferometry circuit diagram.} The quantum circuit used for the interferometry method for measuring string order parameters. The top ancilla qubit is used to perform a controlled unitary that implements the relevant Pauli string in the string order parameter. Measurement of the ancilla qubit is equivalent to the measurement of the string order parameter. Explicit decomposition into elementary gates is provided in Appendix~\ref{ap: interferometry circuits}. }
    \label{fig:interferometry circuit}
\end{figure}

\subsection{Interferometry Experiment}\label{sec: interferometry}

We now consider an alternative method to measure the string order parameters on the quantum computer. This method is motivated by Ramsey or Mach-Zender type interferometry experiments and is know as a Hadamard test in quantum computing. The circuit diagram is given in Fig.~\ref{fig:interferometry circuit}. The basic idea is to use an ancilla qubit prepared in an equal weight superposition using a hadamard gate. We then use a controlled operation that implements the Pauli string $\hat{S}^O$ associated with our string order parameter if the ancilla is in the $|1\rangle$ state. Finally, we apply a Hadamard gate and measure only this ancilla qubit. The expectation value of the ancilla qubit gives the real part of $\langle \psi | \hat{S}^O | \psi \rangle$, which since the Pauli string are Hermitian is equivalent to our string order parameters. We give more details of this method and a decomposition of the circuit into two-qubit gates in Appendix~\ref{ap: interferometry circuits}.

In Fig.~\ref{fig:interferometry results} we show the results from the IBM quantum computer for the interferometry experiments where we see a significant improvement over the results shown in Fig.~\ref{fig:quant_comp}. The quantitative accuracy is improved for all string lengths and a far reduced dependence on the length is observed. In fact, the measured values do not seem to decrease monotonically with increasing string length. This suggests that the fluctuations between these different runs is comparable or larger than the increased error from the deeper circuit. Importantly the qualitative behaviour, including the location of the phase transition is clearly visible for both string order parameters for all string lengths $l$.

\begin{figure}[t]
	\centering
	\raisebox{145pt}{\textbf{(a)}}
	\hspace{-8pt}
	\begin{tikzpicture}
		\begin{axis}[
			height = 0.65\columnwidth,
			width = 0.98\columnwidth,
			xtick = {-1,-0.75,-0.5,...,1},
			xmin=-1.08,xmax=1.08,
			ytick = {-0.2,-0.1,0,0.1,0.2,0.3,0.4,0.5,0.6,0.7,0.8,0.9,1},
			yticklabels = {-0.2,-0.1,0,0.1,0.2,0.3,0.4,0.5,0.6,0.7,0.8,0.9,1},
			xlabel = Tuning parameter $g$,
			ylabel = String order parameter $\mathcal{S}^{ZY}$,
			grid = major,
			legend pos = north east]
		\addplot[color=gray,line width=1.2pt,domain=-1:0,samples=50]{-1*4*x/(1-x)^2};
		\addlegendentry{exact}
		\addplot[color=gray,line width=1.2pt,domain=0:1,samples=50,forget plot]{0};
		\addplot[green-3, only marks, mark=triangle*, mark size=3]
		table[x = g, y = sop] {data_2021-10-25/real_data_N7_SO1_int_backend_ibmq_toronto_2021-10-25.txt};
		\addlegendentry{$l = 5$}
		\addplot[green-4, only marks, mark=diamond*, mark size=3]
		table[x = g, y = sop] {data_2021-10-25/real_data_N8_SO1_int_backend_ibmq_toronto_2021-10-25.txt};
		\addlegendentry{$l=6$}
		\addplot[green-5, only marks, mark=pentagon*, mark size=2.5]
		table[x = g, y = sop] {data_2021-10-25/real_data_N9_SO1_int_backend_ibmq_toronto_2021-10-25.txt};
		\addlegendentry{$l=7$}
		\end{axis}
	\end{tikzpicture}\\
	\vspace{-3pt}
	\raisebox{145pt}{\textbf{(b)}}
	\hspace{-8pt}
	\begin{tikzpicture}
		\begin{axis}[
			height = 0.65\columnwidth,
			width = 0.98\columnwidth,
			xtick = {-1,-0.75,-0.5,...,1},
			xmin=-1.08,xmax=1.08,
			ytick = {-0.2,-0.1,0,0.1,0.2,0.3,0.4,0.5,0.6,0.7,0.8,0.9,1},
			yticklabels = {-0.2,-0.1,0,0.1,0.2,0.3,0.4,0.5,0.6,0.7,0.8,0.9,1},
			xlabel = Tuning parameter $g$,
			ylabel = String order parameter $\mathcal{S}^\mathbb{1}$,
			grid = major,
			legend pos = north west]
		\addplot[color=gray,line width=1pt,domain=0:1]{1*4*x/(1+x)^2};
		\addlegendentry{exact}
		\addplot[color=gray,line width=1pt,domain=-1:0,forget plot]{0};
		\addplot[green-1, only marks, mark size=2.2]
		table[x = g, y = sop] {data_2021-10-25/real_data_N5_SO2_int_backend_ibmq_toronto_2021-10-25.txt};
		\addlegendentry{$l=3$}
		\addplot[green-2, only marks, mark=square*, mark size=2]
		table[x = g, y = sop] {data_2021-10-25/real_data_N6_SO2_int_backend_ibmq_toronto_2021-10-25.txt};
		\addlegendentry{$l=4$}
		\addplot[green-3, only marks, mark=triangle*, mark size=3]
		table[x = g, y = sop] {data_2021-10-25/real_data_N7_SO2_int_backend_ibmq_toronto_2021-10-25.txt};
		\addlegendentry{$l=5$}
		\addplot[green-4, only marks, mark=diamond*, mark size=3]
		table[x = g, y = sop] {data_2021-10-25/real_data_N8_SO2_int_backend_ibmq_toronto_2021-10-25.txt};
		\addlegendentry{$l=6$}
		\addplot[green-5, only marks, mark=pentagon*, mark size=2.5]
		table[x = g, y = sop] {data_2021-10-25/real_data_N9_SO2_int_backend_ibmq_toronto_2021-10-25.txt};
		\addlegendentry{$l=7$}
		\end{axis}
	\end{tikzpicture}
	\caption{\textbf{Identification of the phase transition using interferometry.} \textbf{(a)} Results of the string order parameter $\mathcal{S}^{ZY}(g)$ of length $l=5,6,7$, simulated using number of qubits $N=7,8,9$, respectively. \textbf{(b)} $\mathcal{S}^{\mathbb{1}}(g)$ of length $l=3,4,5,6,7$, corresponding to $N = 5,6,7,8,9$. We compare with the analytic results.}
	\label{fig:interferometry results}
\end{figure}

\section{Error Modelling}\label{sec: error modelling}

In our results from the IBM quantum computers there was a stark contrast in the quality of results obtained by the two different methods. While one used shallow circuits and measured many qubits, the other measured only one qubit at the cost of deeper circuits. Therefore the two methods are impacted differently by the various sources of errors in these quantum devices. In this section we use simple models to analyse the impact of three different sources of errors: measurement error, unitary gate errors, and decoherence. In Fig.~\ref{fig:error} we plot the relative error in the string order parameter against the length of the string order parameter and the relevant error rate. The relative error is computed at $g=0.5$ for $S^\mathbb{1}$ and $g=-0.5$ for $S^{ZY}$.

\begin{figure*}[t!]
	\centering
	\vspace{0pt}
	\raisebox{105pt}{\textbf{(a)}}
	\hspace{-20pt}
	\begin{tikzpicture}
		\begin{axis}[
		    title = \textbf{Measurement error},
			height = 0.24\textwidth,
			width = 0.35\textwidth,
			xtick = {7,8,9,10,11,12},
			xticklabels = {5,6,7,8,9,10},
			xmin = 6.5 ,xmax = 12.5,
			ytick = {0, 0.1, 0.2, 0.3, 0.4, 0.5},
			yticklabels = {0, 0.1, 0.2, 0.3, 0.4, 0.5},
			xlabel = String length $l$,
			ylabel = Relative error,
			grid = major,
			legend style={nodes={scale=0.85, transform shape}}]
		\addplot[blue-5, mark=*, line width=1.2pt,  mark size=2.2]
		table[x = qubits, y = error] {error_models/measure_qubits_so1.txt};
		\addplot[green-5, mark=square*, line width=1.2pt, mark size=2]
		table[x = qubits, y = error] {error_models/measure_qubits_so1_int.txt};
		\addplot[blue-2, mark=triangle, line width=1.2pt, mark size=3]
		table[x = qubits, y = error] {error_models/measure_qubits_so2.txt};
		\addplot[green-2, mark=diamond, line width=1.2pt, mark size=3]
		table[x = qubits, y = error] {error_models/measure_qubits_so2_int.txt};
		\node[style={fill=white}] at (axis cs: 7.5,.45) {$\epsilon=0.03$};
		\end{axis}
	\end{tikzpicture}
	\vspace{0pt}
	\raisebox{105pt}{\textbf{(c)}}
	\hspace{-20pt}
	\begin{tikzpicture}
		\begin{axis}[
		    title = \textbf{Unitary gate errors},
			height = 0.24\textwidth,
			width = 0.35\textwidth,
			xtick = {7,8,9,10,11,12},
			xticklabels = {5,6,7,8,9,10},
			xmin = 6.5 ,xmax = 12.5,
			xlabel = String length $l$,
			ylabel = Relative error,
			grid = major]
		\addplot[blue-5, line width=1.2pt, mark=*, mark size=2.2]
		table[x = qubits, y = error] {error_models/gate_avg_100_qubits_so1.txt};
		\addplot[green-5, line width=1.2pt, mark=square*, mark size=2]
		table[x = qubits, y = error] {error_models/gate_avg_100_qubits_so1_int.txt};
		\addplot[blue-2, line width=1.2pt, mark=triangle, mark size=3]
		table[x = qubits, y = error] {error_models/gate_avg_100_qubits_so2.txt};
		\addplot[green-2, line width=1.2pt, mark=diamond, mark size=3]
		table[x = qubits, y = error] {error_models/gate_avg_100_qubits_so2_int.txt};
		\node[style={fill=white}] at (axis cs: 7.5,.31) {$\epsilon=0.05$};
		\end{axis}
	\end{tikzpicture}
	\vspace{0pt}
	\raisebox{105pt}{\textbf{(e)}}
	\hspace{-20pt}
	\begin{tikzpicture}
		\begin{axis}[
		    title = \textbf{Decoherence},
			height = 0.24\textwidth,
			width = 0.35\textwidth,
			xtick = {7,8,9,10,11,12},
			xticklabels = {5,6,7,8,9,10},
			xmin = 6.5 , xmax = 12.5,
			ytick = {0, 0.06,0.08,0.1},
			yticklabels = {0, 0.06,0.08,0.1},
			xlabel = String length $l$,
			ylabel = Relative error,
			grid = major]
		\addplot[blue-5, line width=1.2pt, mark=*, mark size=2.2]
		table[x = qubits, y = error] {error_models/depolarization_qubits_so1.txt};
		\addplot[green-5, line width=1.2pt, mark=square*, mark size=2]
		table[x = qubits, y = error] {error_models/depolarization_qubits_so1_int.txt};
		\addplot[blue-2, line width=1.2pt, mark=triangle, mark size=3]
		table[x = qubits, y = error] {error_models/depolarization_qubits_so2.txt};
		\addplot[green-2, line width=1.2pt, mark=diamond, mark size=3]
		table[x = qubits, y = error] {error_models/depolarization_qubits_so2_int.txt};
		\node[style={fill=white}] at (axis cs: 7.7,.095) {$T=100\mu s$};
		\end{axis}
	\end{tikzpicture}
	
	\vspace{0pt}
	\raisebox{105pt}{\textbf{(b)}}
	\hspace{-20pt}
	\begin{tikzpicture}
		\begin{axis}[
			height = 0.24\textwidth,
			width = 0.35\textwidth,
			xticklabels = {0,0.02,0.04,0.06},
			xtick = {0,0.02,0.04,0.06},
			ytick = {-0.2,-0.1,0,0.1,0.2,0.3,0.4,0.5,0.6,0.7,0.8,0.9,1},
			xlabel = Bit flip probability,
			ylabel = Relative error,
			scaled x ticks = false,
			grid = major]
		\addplot[blue-5, line width=1.2pt, mark=*, mark size=2.2]
		table[x = rate, y = error] {error_models/measure_rate_so1.txt};
		\addplot[green-5, line width=1.2pt, mark=square*, mark size=2]
		table[x = rate, y = error] {error_models/measure_rate_so1_int.txt};
		\addplot[blue-2, line width=1.2pt,  mark=triangle, mark size=3]
		table[x = rate, y = error] {error_models/measure_rate_so2.txt};
		\addplot[green-2, line width=1.2pt, mark=diamond, mark size=3]
		table[x = rate, y = error] {error_models/measure_rate_so2_int.txt};
		\node[style={fill=white}] at (axis cs: 0.005,.45) {$l=5$};
		\end{axis}
	\end{tikzpicture}
	\vspace{0pt}
	\raisebox{105pt}{\textbf{(d)}}
	\hspace{-20pt}
	\begin{tikzpicture}
		\begin{axis}[
			height = 0.24\textwidth,
			width = 0.35\textwidth,
			xtick = {0,0.02,0.04,0.06,0.08,0.1},
			xticklabels = {0,0.02,0.04,0.06,0.08,0.1},
			xmin=-0.01 ,xmax=0.11,
			ytick = {-0.2,-0.1,0,0.1,0.2,0.3,0.4,0.5,0.6,0.7,0.8,0.9,1},
			xlabel = Error rate,
			ylabel = Relative error,
			grid = major]
		\addplot[blue-5, line width=1.2pt, mark=*, mark size=2.2]
		table[x = rate, y = error] {error_models/gate_avg_100_rate_so1.txt};
		\addplot[green-5, line width=1.2pt, mark=square*, mark size=2]
		table[x = rate, y = error] {error_models/gate_avg_100_rate_so1_int.txt};
		\addplot[blue-2, line width=1.2pt, mark=triangle, mark size=3]
		table[x = rate, y = error] {error_models/gate_avg_100_rate_so2.txt};
		\addplot[green-2, line width=1.2pt, mark=diamond, mark size=3]
		table[x = rate, y = error] {error_models/gate_avg_100_rate_so2_int.txt};
		\end{axis}
	\end{tikzpicture}
	\vspace{0pt}
	\raisebox{105pt}{\textbf{(f)}}
	\hspace{-20pt}
	\begin{tikzpicture}
		\begin{axis}[
			height = 0.24\textwidth,
			width = 0.35\textwidth,
			xtick = {20,50,80,...,200},
			xmin=10 ,xmax=210,
			ytick = {0,0.05,0.1,0.15,0.2,0.25},
			yticklabels = {0,0.05,0.1,0.15,0.2,0.25},
			xlabel = Decoherence time $T$ ($\mu$s),
			ylabel = Relative error,
			grid = major,
			legend pos = north east]
		\addplot[blue-5, line width=1.2pt, mark=*, mark size=2.2]
		table[x = T, y = error] {error_models/depolarization_rate_so1.txt};
		\addlegendentry{$S^{ZY}$}
		\addplot[green-5, line width=1.2pt, mark=square*, mark size=2]
		table[x = T, y = error] {error_models/depolarization_rate_so1_int.txt};
		\addlegendentry{$S^{ZY}$ int.}
		\addplot[blue-2, line width=1.2pt, mark=triangle, mark size=3]
		table[x = T, y = error] {error_models/depolarization_rate_so2.txt};
		\addlegendentry{$S^\mathbb{1}$}
		\addplot[green-2, line width=1.2pt, mark=diamond, mark size=3]
		table[x = T, y = error] {error_models/depolarization_rate_so2_int.txt};
		\addlegendentry{$S^\mathbb{1}$ int.}
		\end{axis}
	\end{tikzpicture}
	
	\caption{\textbf{Relative error for simple error models.} Numerical simulations for three simple error models: \textbf{(a-b)} Measurement errors as described in Sec.~\ref{sec: measurement error}; \textbf{(c-d)} Unitary gate errors as described in Sec.~\ref{sec: gate error}; \textbf{(e-f)} Decoherence due to polarizing channel as described in Sec.~\ref{sec: decoherence}. Subfigures \textbf{(a)}, \textbf{(c)}, \textbf{(e)} show the dependence of the relative error on string length with bit flip probability 0.03, error rate 0.05, and decoherence time 100$\mu$s, respectively. Subfigures \textbf{(b)}, \textbf{(d)}, \textbf{(f)} show the dependence on the relevant error rate for the model for string length $l=5$. Solid symbols correspond to the relative error of $S^{ZY}(g=-0.5)$, and hollow symbols to $S^\mathbb{1}(g=0.5)$. Data simulated using the interferometry method is labelled "int." in the legend. }
	\label{fig:error}
\end{figure*}

\subsection{Measurement Error}\label{sec: measurement error}

First, let us consider the measurement error. We use a simple model for independent bit flip errors, where during the measurement process there is a probability $\epsilon\in[0,1]$ that a qubit is flipped. To simulate this process we consider the pure state $|\psi\rangle$ produced by the circuit and construct a vector containing the probabilities of measuring each bit string, i.e., $p_i = |\langle i | \psi\rangle|^2$, where $|i\rangle$ is a computational basis state. We then construct an error matrix $M_\epsilon$, whose elements are given by
\begin{equation}
    \langle i | M_\epsilon | j \rangle = \epsilon^{N_{ij}} (1-\epsilon)^{N-N_{ij}},
\end{equation}
where $N$ is the total number of qubits and $N_{ij}$ is the number of bit flips between $|i\rangle$ and $|j\rangle$. As an explicit example, let $|i\rangle = |010011\rangle$ and $|j\rangle = |110010\rangle$, then $N=6$ and $N_{ij} = 2$. We then apply this matrix to our probability vector to get the new probabilities taking into account measurement error, $\tilde{p}_i = \sum_{j} [M_\epsilon]_{ij} p_j$. Finally, we can evaluate the string order parameters from these updated probabilities $\tilde{p}_i$. Despite the simplicity of this measurement error model, a similar model is routinely used to perform effective error mitigation and used by ourselves as explained in Appendix~\ref{ap: error mitigation}.

The results from the measurement error model are shown in Figs.~\ref{fig:error}(a) and (b). The first thing to note is that the relative error of the direct method increases with the length of the string order parameter, but doesn't for the interferometry method. This is because in the latter we are only ever measuring a single qubit, and so the probability of a measurement error is simply given by $\epsilon$. In contrast, when measuring multiple qubits, errors in any of these qubits affects the string order parameter. Furthermore, this means the direct method is also significantly more sensitive to increasing the probability of bit flip errors, as shown in Fig.~\ref{fig:error}(b). For the IBM device that we used, the average readout error corresponded to approximately $\epsilon = 0.03$. Our simulations show that this alone could account for approximately a 30\% error in the string order parameter for $l=5$ for the direct method, with even larger errors for $l>5$.

\subsection{Unitary Gate Error}\label{sec: gate error}

The second type of error that we will consider is unitary gate errors. These might correspond to imperfectly calibrated gates or drift in the device that means the intended unitary is not implemented perfectly. For this simulation we assume that single qubit gates are implemented perfectly, but the entangling CNOT gates are subject to random unitary perturbations. This will be done in the following way,
\begin{equation}
    \text{CNOT} \rightarrow \text{exp}\left[\text{log}(\text{CNOT}) + i \epsilon H\right],
\end{equation}
where $H=(M + M^\dag) / 2$ and $M$ has complex matrix elements drawn from a normal random distribution, and $\epsilon$ controls the error rate. Here $\text{exp}$ and $\text{log}$ refer to matrix exponential and logarithm, respectively. In our simulations we average over 100 realizations of the random gates.

As shown, in Fig.~\ref{fig:error}(c), the error for both methods now increases as a function of the string length. Furthermore, the interferometry method is now more sensitive to this type of error due to the increased number of gates in the circuit. However, the difference between the two is much less drastic than in the case of measurement errors. Unfortunately, we are not able to quantify the amount of unitary gate error in the IBM quantum computers since it is difficult to disentangle unitary gate errors from other decoherent sources of error.

\subsection{Decoherence}\label{sec: decoherence}

Despite the remarkable amount of isolation achieved to realise current quantum computers, these devices are not perfectly closed systems. As a result, the qubits interact with the environment and will be eventually become decoherent. There are many forms of decoherence that happen in a realistic device but for simplicity we focus only on the depolarizing channel. Furthermore, while the decoherence is in reality happening throughout the implementation of the circuit, as well as state preparation and readout, we will apply the depolarizing channel at once, immediately prior to measurement. While this is certainly an approximation of reality, this model has also been used to effectively mitigate errors on quantum computers~\cite{Vovrosh2021}, which verifies its approximate validity on current devices.

Concretely, following the application of our quantum circuit we will be in the pure state $|\psi\rangle$ and then the depolarizing channel will result in the mixed state represented by the density matrix
\begin{equation}
    \rho = \epsilon\frac{\mathbb{1}}{2^N} + (1-\epsilon) |\psi \rangle \langle \psi |.
\end{equation}
The parameter $\epsilon$ controls the amount to which the pure state is mixed with a completely mixed density matrix. The parameter $\epsilon = (1-e^{-t/T})$, where $t$ is the total time of the circuit and $T$ is the time scale for decoherence. Due to an order of magnitude difference in time scales, we assume single qubit gates are instantaneous whereas CNOT gates are implemented in 425ns, the average for the IBM quantum computer we used. The total time is then given by the gate time multiplied by the number of asynchronous CNOT gates. The corresponding decoherence time for the device was approximately $100\mu$s.

The error due to this depolarizing channel appears to scale approximately linearly with the string length $l$, as shown in Fig.~\ref{fig:error}(e). This corresponds to the linear increase in circuit depth with $l$. However, the difference between the direct and interferometry methods is small. This is because these methods only differ by two asynchronous layers of CNOTS. The rest of the additional CNOT gates can be done in parallel with the gates common to both methods. Additionally, the circuit for $S^{\mathbb{1}}$ measured at $g=0.5$ has one additional CNOT gate compared with $S^{ZX}$ at $g=-0.5$, the effect of which can also be seen in this figure. In Fig.~\ref{fig:error}(f) we plot the relative error as a function of the decoherence time, which shows a characteristic exponential behaviour.

\section{Discussion}\label{sec: discussion}
Above we have focused on a particular line through the phase diagram in Fig.~\ref{fig:phase_diagram}, which has an especially efficient construction of the ground states. This enabled an exact representation within the limitations of existing devices. In this paper we have considered a particularly simple path, but our approach is general and potentially provides a genuine advantage to using NISQ devices. In fact, all matrix product states can be constructed in a similar way~\cite{Schon2005a,Barratt2020} and can be variationally optimised on a quantum computer~\cite{Barratt2020}. Such variational solvers have already been demonstrated in the setting of small molecules~\cite{Kandala2017,OMalley2016} using variational quantum eigensolvers (VQE)~\cite{Peruzzo2014}.


It is still an open and interesting problem to find optimal ansatz circuits for variational optimization. A recent work has shown that sequential quantum circuit ans{\"a}tze---similar to the ones used in this paper---are efficient "sparse" representations for some quantum ground states and in simulating non-equilibrium dynamics~\cite{lin2020}. By directly using the connectivity of the quantum computers it may be possible to go beyond what is accessible with classical numerics in two-dimensions with shallow depth (polynomial in system size) quantum circuits. In particular, it is often numerically expensive to compute correlators in higher-dimensional tensor networks. Representing these as quantum circuits~\cite{Banuls2008} will permit considerable speedup in their manipulation and measurement---with a potential exponential advantage in certain circumstances. As a concrete example, there exists a simple representation of topologically ordered string-net models~\cite{Levin2005} in terms of tensor networks~\cite{Buerschaper2009,Gu2009}, that nevertheless remains difficult to deal with numerically. 

We demonstrated experimentally on the IBM quantum computers two alternative methods for measuring non-local string order parameters. While direct measurement of multiple qubits was highly sensitive to increased string length, the interferometric method gave consistent qualitative agreement to the exact results. By analysing three simple error models we identified readout errors as those that dominated our results, and so strongly favoured the interferometric approach which required measurement of a single qubit. It is possible that in other quantum computer technologies, the extent of measurement and gate errors (either coherent or decoherent) would be reversed and in that case the direct method might be preferred. It is clear that in the NISQ era it is important to understand the sources of error of a particular device and tailor our approach accordingly.



Beyond SPT phases, where we know how to construct the order parameters, we need to find efficient ways of detecting and differentiating different phases. Recent work proposes quantum-hybrid algorithms based on ideas from machine learning and renormalization group~\cite{Grant2018,Cong2019}. These algorithms are scalable and practical to implement on near-term devices. The combination of machine learning tools and quantum hardware is potentially very powerful with many applications~\cite{Biamonte2017}.

In this paper we have distinguished two topologically inequivalent phases and identified the transition between them using a real quantum device. We compared two complimentary methods and analysed their relative accuracy using simple error models. Despite the infancy of the current technology, our work clearly demonstrates that near-term NISQ devices can be used as practical tools for the study of condensed matter physics.

\begin{acknowledgements}
	\noindent We thank Julian Bibo and Ruben Verresen for helpful discussions. Quantum circuit diagrams were produced using the Quantikz latex package~\cite{quantikz}. We acknowledge the Samsung Advanced Institute of Technology Global Research Partnership and the Research Institute CODE of the Universit{\"a}t der Bundeswehr M{\"u}nchen for providing access to the IBM Q quantum computer. A.S. was supported by a Research Fellowship from the Royal Commission for the Exhibition of 1851. A.S. and F.P. were in part funded by the European Research Council (ERC) under the European Union's Horizon 2020 research and innovation programme (grant agreement No. 771537). A.G.G was supported by the EPSRC. B.J. was supported by a scholarship from the Hanns-Seidel-Stiftung.
\end{acknowledgements}

\appendix

\section{Connection to infinite MPS}\label{ap: MPS}

Using matrix product states it is possible to represent infinite translationally invariant states with local entanglement~\cite{Vidal2007}. This representation consists of a number of finite bond dimension tensors in a unit cell---that are in principal repeated infinitely many times---and the eigenvectors of the corresponding transfer matrix. The left and right eigenvectors of the transfer matrix allow us to compute expectation values with a finite cost by terminating the tensor contraction beyond the support of our observable, similar to our termination of the quantum circuit.

The ground states along the path in the main text have an infinite MPS representation with single site unit cell and bond dimension 2~\cite{Wolf2006}. The state is therefore described by a single 3 index tensor $M^i_{ab}$, where $i = \uparrow, \downarrow$ is the physical index and $a,b = 0,1$ are the virtual indices with bond dimension 2. To relate this to a quantum circuit we must first transform the tensors to right canonical form (equivalently left canonical). Right canonical form amounts to the defining new tensors $B^i_{ab} = \sum_{cd} X_{ac} M^i_{cd} X^{-1}_{d,b}$ for some invertible matrix $X$ such that $\sum_{i b} B^i_{ab} [B^{i}_{cb}]^* = \delta_{a,c}$. We can represent this type of tensor using a 2-qubit unitary gate, which we write as $U^{ij}_{kl} = U_{(ij),(kl)}$, where $i,k$ refer to the first qubit and $j,l$ to the second. More explicitly, $\hat{U} = \sum_{ijkl} U^{ij}_{kl}|i,j \rangle \langle k,l|$. The tensors are then given by the matrix elements $B^{i}_{ab} = U^{ib}_{a \uparrow}$, and the condition of right canonical form is equivalent to the unitarity of $U$. This allows us to construct the circuit shown in the main text but the connection to infinite MPS is more general and not restricted to bond dimension 2~\cite{Barratt2020}. Infinite MPS have been an extremely successful approach since they avoid boundary effects that are present for finite systems. In this paper we are able to translate the success of these methods to existing quantum computers.

\subsection{Quantum circuit in elementary gates}\label{ap: circuit decomposition}

In this section we give the details of the quantum circuit shown schematically in Fig.~\ref{fig:circuit_structure} of the main text. We further decompose these circuits into the native gates that can be implemented on the IBM Q devices. The native gates are the single qubit rotations
\begin{equation}
\begin{quantikz}[row sep=0.1cm, column sep=0.2cm]
		 & \gate{U_3(\theta,\phi,\lambda)} & \qw
	\end{quantikz}	
=
\left(\begin{array}{cc}
\cos\frac{\theta}{2} & -e^{i\lambda}\sin\frac{\theta}{2}\\
e^{i\phi}\sin\frac{\theta}{2} & e^{i(\lambda+\phi)}\cos\frac{\theta}{2}
\end{array} \right).
\end{equation}
and the controlled-not (CNOT) entangling operation.
\begin{equation}
\begin{quantikz}[row sep=0.5cm, column sep=0.2cm]
		 & \ctrl{1} & \qw \\
		 & \targ{} & \qw 
	\end{quantikz} \, = \, \left(\begin{array}{cccc}
1 & 0 & 0 & 0 \\
0 & 1 & 0 & 0 \\
0 & 0 & 0 & 1 \\
0 & 0 & 1 & 0
\end{array}\right).
\end{equation}
We will also use two special single qubit gates
\begin{equation}
\begin{quantikz}[row sep=0.1cm, column sep=0.2cm]
		 & \gate{X} & \qw
	\end{quantikz}	= 
\left(\begin{array}{cc}
0 & 1 \\
1 & 0
\end{array} \right), \quad
\begin{quantikz}[row sep=0.1cm, column sep=0.2cm]
		 & \gate{H} & \qw
	\end{quantikz}	= 
\frac{1}{\sqrt{2}}\left(\begin{array}{cc}
1 & 1 \\
1 & -1
\end{array} \right),
\end{equation}
the Pauli-X gate and the Hadamard gate. 

In order to construct the quantum circuits for the states along the path we consider, we first start with an MPS representation of this path. This representation was given in Ref.~\cite{Wolf2006} and consists of two matrices for each site of the chain
\begin{equation}
    M^{\uparrow} = \left( 
    \begin{array}{cc}
    0 & 0 \\
    1 & 1
    \end{array}
    \right), \quad
    M^{\downarrow} = \left( 
    \begin{array}{cc}
    1 & g \\
    0 & 0
    \end{array}
    \right),
\end{equation}
for $g\in[-1,1]$. We next have to put this MPS representation into either left or right canonical form. In right canonical form the matrices become
\begin{equation}
\begin{aligned}
    B^{\uparrow} &= \frac{1}{\sqrt{1+|g|}}\left( 
    \begin{array}{cc}
    0 & 0 \\
    \sqrt{|g|} & 1
    \end{array}
    \right),\\
    B^{\downarrow} &= \frac{1}{\sqrt{1+|g|}} \left( 
    \begin{array}{cc}
    1 & \text{sign}(g)\sqrt{|g|} \\
    0 & 0
    \end{array}
    \right).
\end{aligned}
\end{equation}
In this form the dominant left eigenvector (with eigenvalue 1) of the transfer matrix $T_{(\alpha\alpha'),(\beta\beta')} = \sum_{j=\uparrow,\downarrow} B^{j}_{\beta,\alpha} B^{j}_{\beta', \alpha'}$ corresponds to the identity. This eigenvalue equation in terms of the unitary gates $U$ and $U_1$ is shown in Fig.~\ref{fig:eigen}. The dominant right eigenvector (with eigenvalue 1) can be written as $V_{(\beta,\beta')} = \sum_{j=\uparrow,\downarrow} B^{[1]j}_{\beta} B^{[1]j}_{\beta'}$, where
\begin{equation}
\begin{aligned}
    B^{[1]\uparrow} &= \frac{1}{\sqrt{2(1+|g|)}}\left( 
    \begin{array}{cc}
    \sqrt{|g|}, & 1
    \end{array}
    \right),\\
    B^{[1]\downarrow} &= \frac{1}{\sqrt{2(1+|g|)}} \left( 
    \begin{array}{cc}
    1, & \text{sign}(g)\sqrt{|g|}
    \end{array}
    \right).
\end{aligned}
\end{equation}

\begin{figure}[t]
	\centering
\begin{quantikz}[row sep=0.1cm, column sep=0.1cm]
		\lstick{$|0\rangle$} & \gate[style={fill=nice-green!20},wires=2][0.5cm]{U_1} & \qw & \qw & \qw & \qw & \qw & \qw & \qw & \gate[style={fill=nice-green!20},wires=2][0.5cm]{U_1} & \qw & \rstick{$|0\rangle$} \\
		\lstick{$|0\rangle$} & & \gate[style={fill=nice-blue!20},wires=2][0.5cm]{U} & \qw & \qw & \qw & \qw & \qw & \gate[style={fill=nice-blue!20},wires=2][0.5cm]{U} & & \qw &\rstick{$|0\rangle$}  \\
		\lstick{$|0\rangle$} & \qw & & \qw & \beta & \qquad & \beta' & \, & \qw & \qw & \qw & \rstick{$|0\rangle$}  \\
\end{quantikz}

= \,\begin{quantikz}[row sep=0.1cm, column sep=0.1cm]
		\lstick{$|0\rangle$} & \gate[style={fill=nice-green!20},wires=2][0.5cm]{U_1} & \qw & \qw & \qw & \qw & \qw & \gate[style={fill=nice-green!20},wires=2][0.5cm]{U_1} & \qw &\rstick{$|0\rangle$}  \\
		\lstick{$|0\rangle$} & & \qw & \beta & \qquad & \beta' & \, & \qw & \qw & \rstick{$|0\rangle$}  \\
\end{quantikz}
	\caption{\textbf{Left eigenvector of the transfer matrix.} Given the repeating unitary gate $U$, the unitary gate $U_1$ is implicitly defined by this left eigenproblem, with eigenvalue $1$. The solution to this equation is given in Eq.~\eqref{eq: U matrices}.  }\label{fig:eigen}
\end{figure}

Given this canonical form, we can then embed these matrices inside two-qubit unitaries as follows
\begin{equation}\label{eq: U matrices}
\begin{aligned}
    U = \frac{1}{\sqrt{1+|g|}} \left(
    \begin{array}{cccc}
    {\color{red} 0} & \times & {\color{red} \sqrt{|g|}} & \times \\
    {\color{red} 0} & \times & {\color{red} 1} & \times \\
    {\color{blue} 1} & \times & {\color{blue} 0} & \times \\
    {\color{blue} \text{sign}(g)\sqrt{|g|}} & \times & {\color{blue} 0} & \times
    \end{array}
    \right), \\
    U_1 = \frac{1}{\sqrt{2(1+|g|)}} \left(
    \begin{array}{cccc}
    {\color{red} \sqrt{|g|}} & \times & \times & \times \\
    {\color{red} 1} & \times & \times & \times \\
    {\color{blue} 1} & \times & \times & \times \\
    {\color{blue} \text{sign}(g)\sqrt{|g|}} & \times & \times & \times
    \end{array}
    \right),
\end{aligned}
\end{equation}
where crosses mark elements of the unitary that we are free to choose up to the unitarity constraint. The elements coloured red correspond to the transpose of $B^{\uparrow}$ (resp. $B^{[1]\uparrow}$) and those coloured blue to the transpose of $B^{\downarrow}$ (resp. $B^{[1]\downarrow}$). The $U_1$ unitary gates satisfy the left eigenvalue equation shown in Fig.~\ref{fig:eigen}. Finally, the gate sequences shown in Fig.~\ref{fig:circuit_structure} and their angles were found by inspection such that their matrix representations match those in Eq.~\eqref{eq: U matrices}.

The quantum circuits in Fig.~\ref{fig:circuit_structure} contain three two-qubit gates that need to be decomposed further into the elementary gate set. These are all of the form of controlled unitary gates. The first is the controlled-Z or controlled-phase gate
\begin{equation}
\raisebox{2pt}{\begin{quantikz}[row sep=0.6cm, column sep=0.2cm]
		 & \ctrl{1} & \qw \\
		 & \control{} & \qw 
	\end{quantikz}} \, = \, \begin{quantikz}[row sep=0.4cm, column sep=0.2cm]
		 & \qw & \ctrl{1} & \qw & \qw \\
		 & \gate{H} & \targ{} & \gate{H} &\qw 
	\end{quantikz}.
\end{equation}
The other two are of the form
\begin{equation}
\begin{quantikz}[row sep=0.5cm, column sep=0.2cm]
		 & \ctrl{1} & \qw \\
		 & \gate{V} & \qw 
\end{quantikz}, \quad \text{and} \quad
\begin{quantikz}[row sep=0.5cm, column sep=0.2cm]
		 & \octrl{1} & \qw \\
		 & \gate{W} & \qw 
	\end{quantikz} = \raisebox{2pt}{\begin{quantikz}[row sep=0.4cm, column sep=0.2cm]
		 & \gate{X} & \ctrl{1} & \gate{X} & \qw \\
		 & \qw & \gate{W} & \qw &\qw 
	\end{quantikz}},
\end{equation}
where the single qubit gates are of the form
\begin{equation}
V = 
\left( \begin{array}{cc}
\sin \theta_v & \cos \theta_v \\
\cos \theta_v & -\sin \theta_v
\end{array}\right),
\end{equation}
and similarly for $W$. For these single qubit gates the controlled-unitary gate can be implemented using a single CNOT as follows
\begin{equation}
    \begin{quantikz}[row sep=0.5cm, column sep=0.2cm]
		 & \ctrl{1} & \qw \\
		 & \gate{V} & \qw 
	\end{quantikz} = \raisebox{0pt}{\begin{quantikz}[row sep=0.5cm, column sep=0.2cm]
		 & \qw & \ctrl{1} & \qw & \qw \\
		 & \gate{\tilde{V}} & \targ{} & \gate{\tilde{V}^\dag} &\qw,
	\end{quantikz}}
\end{equation}
where $\tilde{V} = U_3(\theta_v,0,0)$ and $\tilde{W} = U_3(\theta_w,0,0)$, and $R = U_3(\theta_r,0,\pi)$ for the gate in Fig.~\ref{fig:circuit_structure}(U1), with angles specified by
\begin{equation}
\begin{aligned}
\theta_v &= \arcsin\left(\frac{\sqrt{|g|}}{\sqrt{1+|g|}}\right), \quad \theta_v \in [-\pi/2,\pi/2],\\
\theta_w &= \arccos\left(\frac{\text{sign}(g)\sqrt{|g|}}{\sqrt{1+|g|}}\right), \quad \theta_w \in [0,\pi], \\
\theta_r &= 2 \arcsin\left(\frac{1}{\sqrt{1+|g|}}\right), \quad \theta_r \in [-\pi,\pi].
\end{aligned}
\end{equation}
The fully decomposed gates are then shown in Fig.~\ref{fig:gate-decomp}.

\begin{figure}[t]
	\centering
	\hspace{10pt}\raisebox{30pt}{\textbf{(a)}}
	\hspace{-15pt}
	\begin{quantikz}[row sep=0.25cm, column sep=0.2cm]
		 & \gate{X} & \ctrl{1} & \gate{X} & \qw & \ctrl{1} & \gate{X} &  \qw\\
		\lstick{\ket{0}} & \gate{\tilde{W}} & \targ{} & \gate{\tilde{W}^\dag} & \gate{\tilde{V}} & \targ{} & \gate{\tilde{V}^\dag} & \qw & \qw
	\end{quantikz}\hfill\\
	\hspace{10pt}\raisebox{30pt}{\textbf{(b)}}
	\hspace{-15pt}
	\begin{quantikz}[row sep=0.25cm, column sep=0.2cm]
		\lstick{\ket{0}} & \gate{H}  & \ctrl{1} & \qw & \qw & \qw   & \qw\gategroup[wires=2,steps
		=3,style={dashed, rounded corners,fill=gray!30}, background]{Only for $g>0$} & \ctrl{1} & \qw & \qw & \qw\\
		\lstick{\ket{0}} & \qw & \targ{} & \gate{R} & \qw & \qw & \gate{H} & \targ{} & \gate{H} & \qw & \qw \qw
	\end{quantikz}
	\caption{\textbf{Decomposition of the two-qubit gates into the elementary gate set.} \textbf{(a)} The quantum circuit of the two-qubit gate $U$, and \textbf{(b)} The quantum circuit of the two-qubit gate $U1$, shown in Fig.~\ref{fig:circuit_structure} in the main text.}\label{fig:gate-decomp}
\end{figure}

\begin{figure*}[t]
\centering
	\hspace{0pt}\raisebox{70pt}{\textbf{(a)}}
	\hspace{-15pt}
\begin{quantikz}[row sep=0.15cm, column sep=0.10cm]
	    \lstick{\ket{0}} & \gate{H} & \swap{1} & \qw & \qw & \qw & \qw & \qw & \qw & \qw & \qw & \qw & \qw & \qw & \qw & \\
		\lstick{\ket{0}} & \gate[style={fill=nice-green!20},wires=2]{} & \targX{} & \qw & \ctrl{1} & \swap{1} & \qw & \qw & \qw & \qw & \qw & \qw & \qw & \qw & \qw & \\
		\lstick{\ket{0}} & & \gate[style={fill=nice-blue!20},wires=2]{} & \gate[style={fill=nice-red!20}]{H} & \targ{} & \targX{} & \ctrl{1} & \swap{1} & \qw & \qw & \qw & \qw & \qw & \qw & \qw &\\
		\lstick{\ket{0}} & \qw & & \qw & \gate[style={fill=nice-blue!20},wires=2]{} & \gate[style={fill=nice-red!20}]{S^\dag} & \targ{} & \targX{} & \ctrl{1} & \swap{1} & \qw & \qw & \qw & \qw & \qw &\\
		\lstick{\ket{0}} & \qw & \qw & \qw & & \qw & \gate[style={fill=nice-blue!20},wires=2]{} & \qw &  \targ{} & \targX{} & \ctrl{1} & \swap{1} & \qw & \qw & \qw & \\
		\lstick{\ket{0}} & \qw & \qw & \qw & \qw & \qw & \qw & \qw &  \gate[style={fill=nice-blue!20},wires=2]{} & \gate[style={fill=nice-red!20}]{S^\dag} & \targ{} & \targX{} & \ctrl{1} & \gate{H} & \meter{} &\\
		\lstick{\ket{0}} & \qw & \qw & \qw & \qw & \qw & \qw & \qw & & \qw & \gate[style={fill=nice-blue!20},wires=2]{} & \gate[style={fill=nice-red!20}]{H} & \targ{} & \qw & \qw &\\
		\lstick{\ket{0}} & \qw & \qw & \qw & \qw & \qw & \qw & \qw & \qw & \qw & \qw & \qw & \qw & \qw & \qw & \\
\end{quantikz}
\hspace{0pt}\raisebox{70pt}{\textbf{(b)}}
	\hspace{-15pt}
\begin{quantikz}[row sep=0.15cm, column sep=0.10cm]
	    \lstick{\ket{0}} & \gate{H} & \swap{1} & \qw & \qw & \qw & \qw & \qw & \qw & \qw & \qw & \qw & \qw & \qw & \qw & \\
		\lstick{\ket{0}} & \gate[style={fill=nice-green!20},wires=2]{} & \targX{} & \qw & \targ{} & \ctrl{1} & \qw & \qw & \qw & \qw & \qw & \qw & \qw & \qw & \qw & \\
		\lstick{\ket{0}} & & \gate[style={fill=nice-blue!20},wires=2]{} & \gate[style={fill=nice-red!20}]{H} & \ctrl{-1} & \targ{} & \targ{} & \ctrl{1} & \qw & \qw & \qw & \qw & \qw & \qw & \qw &\\
		\lstick{\ket{0}} & \qw & & \qw & \gate[style={fill=nice-blue!20},wires=2]{} & \gate[style={fill=nice-red!20}]{S^\dag} & \ctrl{-1} & \targ{} & \targ{} & \ctrl{1} & \qw & \qw & \qw & \qw & \qw &\\
		\lstick{\ket{0}} & \qw & \qw & \qw & & \qw & \gate[style={fill=nice-blue!20},wires=2]{} & \qw &  \ctrl{-1} & \targ{} & \targ{} & \ctrl{1} & \qw & \qw & \qw & \\
		\lstick{\ket{0}} & \qw & \qw & \qw & \qw & \qw & \qw & \qw &  \gate[style={fill=nice-blue!20},wires=2]{} & \gate[style={fill=nice-red!20}]{S^\dag} & \ctrl{-1} & \targ{} & \ctrl{1} & \gate{H} & \meter{} &\\
		\lstick{\ket{0}} & \qw & \qw & \qw & \qw & \qw & \qw & \qw & & \qw & \gate[style={fill=nice-blue!20},wires=2]{} & \gate[style={fill=nice-red!20}]{H} & \targ{} & \qw & \qw &\\
		\lstick{\ket{0}} & \qw & \qw & \qw & \qw & \qw & \qw & \qw & \qw & \qw & \qw & \qw & \qw & \qw & \qw & \\
\end{quantikz}.
\caption{\textbf{Explicit quantum circuits for the interferometry method.} Circuits shown compute the string order parameter $S^{ZY}$ for $l=5$. \textbf{(a)} Shows the circuit using swap gates to avoid the use of non-local CNOT gates. \textbf{(b)} Using additional simplifications from combing CNOT and swap gates. The greeen and blue gates are $U_1$ and $U$ form the sequential ground state circuit defined in Appendix~\ref{ap: circuit decomposition}. The red gates perform a basis change for the $S^{ZY}$ string order parameter and do not appear for $S^\mathbb{1}$. The white Hadamard gates are those applied to the ancilla in the Hadamard test circuit in Eq.~\eqref{eq: Hadamard test}.}\label{fig: interferometry explicit}
\end{figure*}

\section{Error mitigation}\label{ap: error mitigation}

Because of the high level of readout error in the current generation of quantum computers, we employ a simple error mitigation technique to reduce their effect on our data. This is achieved by constructing a readout matrix that maps between expected ideal basis states and the actual distribution of measurement outcomes. Constructing such a matrix requires partial tomography to measure a full set of basis states. This technique is therefore not scalable but is accessible for the system sizes that can be accurately simulated on the existing devices.

After extracting the readout matrix, we can mitigate the leading readout errors by applying the inverse (or an appropriate pseudo-inverse) of this matrix to the distribution of measurement outcome obtained from the device. This technique was performed using the error-mitigation software built into qiskit (ignis)~\cite{qiskit}.

\section{Interferometry Circuits}\label{ap: interferometry}

\subsection{Derivation}\label{ap: interferometry derivation}

Our interferometry methods for measuring the string order parameters is based on Ramsey or Mach-Zender type interferometry experiments and is also known as a Hadamard test. This method can be summarised by the following circuit
\begin{equation}\label{eq: Hadamard test}
\begin{quantikz}[row sep=0.15cm, column sep=0.15cm]
	    \lstick{\ket{0}} & \qw & \gate{H} & \qw & \ctrl{1} \qw & \gate{H} & \qw & \meter{} \\
		\lstick{\ket{\psi}} & \qwbundle[alternate]{} & \qwbundle[alternate]{} & \qwbundle[alternate]{} & \gate{U} \qwbundle[alternate]{} & \qwbundle[alternate]{} & \qwbundle[alternate]{} & \qwbundle[alternate]{} \\
\end{quantikz}.
\end{equation}
To understand this circuit let us break it into steps. After each of the gates we will have the following states:
\begin{subequations}
\begin{equation}\label{subeq:H1}
    \frac{1}{\sqrt{2}} (|0\rangle + |1\rangle) \otimes |\psi\rangle,
\end{equation}
\begin{equation}\label{subeq:C-U}
    \frac{1}{\sqrt{2}} \left( |0\rangle \otimes |\psi\rangle + |1\rangle \otimes \hat{U}|\psi\rangle \right),
\end{equation}
\begin{equation}\label{subeq:H2}
    \frac{1}{2} \left( |0\rangle \otimes (1+\hat{U})|\psi\rangle + |1\rangle \otimes (1-\hat{U})|\psi\rangle \right).
\end{equation}
\end{subequations}
After the first Hadamard we will get Eq.~\eqref{subeq:H1}. Then applying the controlled unitary we will end up with Eq.~\eqref{subeq:C-U}. Finally, the final Hadamard gives Eq.~\eqref{subeq:H2}. We then measure the ancilla qubit only and construct the $Z$ expectation value. The probability of the ancilla qubit being in the two states is given by
\begin{equation}
\begin{aligned}
    p(|0\rangle) = \frac{1}{4} \left(2 + \langle\psi| \hat{U}^\dag |\psi\rangle + \langle \psi | \hat{U} |\psi\rangle \right), \\ 
    p(|1\rangle) = \frac{1}{4} \left(2 - \langle\psi| \hat{U}^\dag |\psi\rangle - \langle \psi | \hat{U} |\psi\rangle \right).
\end{aligned}
\end{equation}
The expectation value $\langle Z \rangle$ is then given by $p(|0\rangle) - p(|1\rangle) = \text{re}[\langle \psi | \hat{U} | \psi \rangle]$. Since in our case $\hat{U}$ is a Pauli string and so also Hermitian, i.e., $\hat{U}^\dag = \hat{U}$, then the imaginary part is identically zero, and so the string order parameter is real.

\subsection{Circuit Decomposition}\label{ap: interferometry circuits}

For the string order parameter, the unitary we want the expectation value of in the interferometry experiment is a Pauli string either of the form $\sigma^x \cdots \sigma^x$ or $\sigma^z \sigma^y \sigma^x \cdots \sigma^x \sigma^y \sigma^z$. For the former, the controlled unitary is simply a product of CNOT gates, with the control on the ancill and the target on each of the qubits we want to act on. For the latter, we can do a basis transformation using $H$ and $S$ gates and then do the same. 

In order to respect nearest neighbour connectivity, we use swap gates to move the ancilla through the system before applying each CNOT gates, as demonstrated in Fig.~\ref{fig: interferometry explicit}(a). Since the controlled unitary then has a sequential form similar to the state construction, many of these gates can be performed in parallel. This circuit can be additionally simplified by noting that swap gates are equivalent to the product of three CNOT gates acting in alternating directions. By cancelling pairs of CNOT gates we then arrive at the final circuit shown in Fig.~\ref{fig: interferometry explicit} for the case of $S^{ZY}$ and $l=5$. The circuit has a similar structure for all other string order parameters.

\pgfplotsset{every axis legend/.append style={at={(0.5,0.03)},anchor=south,},}
\begin{figure}[t]
	\centering
	\begin{tikzpicture}
		\begin{axis}[
			height = 0.65\columnwidth,
			width = 0.98\columnwidth,
			xtick = {-1,-0.75,-0.5,...,1},
			xmin=-1.08,xmax=1.08,
			ytick = {-1,-1.5,...,-4},
			xlabel = Tuning parameter $g$,
			ylabel = Energy density $\mathcal{E}$,
			grid = major]
		\addplot[color=gray,line width=1.2pt,domain=-1:1,samples=100]{-1*2*(x^2+1)};
		\addlegendentry{exact}
		\addplot[red-1, line width=1.2pt]
		table[x = g, y = energy] {error_models/energy_dephasing_09.txt};
		\addlegendentry{$p = 0.9$}
		\addplot[red-2, line width=1.2pt]
		table[x = g, y = energy] {error_models/energy_dephasing_085.txt};
		\addlegendentry{$p = 0.85$}
		\addplot[red-3, line width=1.2pt]
		table[x = g, y = energy] {error_models/energy_dephasing_08.txt};
		\addlegendentry{$p = 0.8$}
		\addplot[red-4, line width=1.2pt]
		table[x = g, y = energy] {error_models/energy_dephasing_075.txt};
		\addlegendentry{$p = 0.75$}
		\addplot[red-5, line width=1.2pt]
		table[x = g, y = energy] {error_models/energy_dephasing_07.txt};
		\addlegendentry{$p = 0.7$}
		\end{axis}
	\end{tikzpicture}
	\caption{\textbf{Effect of dephasing on measuring the energy density.} Results of the simple depahsing model defined in Eq.~\eqref{eq: dephasing model} for different values of $p$. These results qualitatively reproduce the experimentally measured data shown in Fig.~\ref{fig:energy}. }
	\label{fig:energy dephasing}
\end{figure}

In our error model simulations we used both the gate count and the total runtime of the circuit as parameters. We assume that single qubit gates are error free and implemented instantaneously and only the two-qubit CNOT gates are counted. In the case of the direct method we simply have that  both the number of CNOT gates and the number of asynchronous CNOT layers is the same and given by $2l+2$ for $g>0$ and $2l+1$ for $g<0$. For the interferometry method, the total number of CNOT gates for $g>0$ is $4l+6$ and $4l+5$ for $g<0$. The total number of asynchronous CNOT layers is $2l+4$ for $g>0$ and $2l+3$ for $g<0$ due to many gates being performed in parallel, which is only two more layers than the corresponding circuits from the direct method.


\section{Dephasing}\label{ap: dephasing}

Here we present a simple model for the dephasing that leads to the change in shape of the energy density seen in Fig.~\ref{fig:energy}. Dephasing or phase flip errors affect the $X$ and $Y$ basis states but not the $Z$ states. In reality this will also be accompanied by bit flip errors that affect the $Z$ basis states, but here we consider a model of pure phase flip errors. In our simplified model we assume that dephasing has the following effect,
\begin{equation}\label{eq: dephasing model}
    \begin{aligned}
    \langle \hat{\sigma}^x_j \rangle &\rightarrow p \,\langle \hat{\sigma}^x_j \rangle, \\
    \langle \hat{\sigma}^z_j\hat{\sigma}^x_{j+1}\hat{\sigma}^z_{j+2} \rangle &\rightarrow p \,\langle \hat{\sigma}^z_j\hat{\sigma}^x_{j+1}\hat{\sigma}^z_{j+2} \rangle, 
    \end{aligned}
\end{equation}
where $p \in [0,1]$, and $\langle \hat{\sigma}^z_j\hat{\sigma}^z_{j+1} \rangle$ is unchanged. The scaled terms are due to the presence of a single Pauli X operator in the expectation value. The scaling parameter $p$ will depend on the rate of dephasing and also the depth of the circuit.

In Fig.~\ref{fig:energy dephasing} we show the effects of this dephasing model on the measured energy density. We see that due to the different scaling of the terms in the Hamiltonian, the shape of the curve changes. This simple model reproduces the qualitative behaviour seen in the experimental data in Fig.~\ref{fig:energy}, and in particular reproduces the dip at $g=0$. Comparing the values of $p$ that we plot with the experimental results, we conclude that there is a significant amount of dephasing in these experiments. 

Note that we did not consider dephasing as a separate error model for the string order parameter because its effect would be similar to the bit flip error, except with a different length dependence between the $S^{ZY}$ and $S^\mathbb{1}$ order parameters. The conclusions made in section~\ref{sec: error modelling} would be unaffected by the inclusion of this simple dephasing model.

\end{document}